# Analysis of Second Harmonic Generation in Photonic-Crystal-Assisted Waveguides

A. D'Orazio[1], D. de Ceglia[1,3], M. De Sario[1]

[1] Dipartimento di Elettrotecnica ed Elettronica, Politecnico di Bari

Via Orabona 4, 70124 Bari, Italy

F. Prudenzano[2]

[2] Dipartimento di Ingegneria dell'Ambiente e per lo Sviluppo Sostenibile,

Politecnico di Bari Viale del Turismo, 8 – Taranto, Italy

M.J. Bloemer[3], M. Scalora[3]

[3] Charles M. Bowden Research Center, AMSRD-AMR-WS-ST, RDECOM

Redstone Arsenal, Alabama 35898-5000, USA

**Abstract**

We study second harmonic generation in a planar dielectric waveguide having a low-index, polymer core layer, bounded by two multilayer stacks. This geometry allows exceptionally strong confinement of the light at the fundamental wavelength inside the core region with virtually zero net propagation losses for distances that exceed several centimeters, provided material and scattering losses are neglected. A phase-matched configuration of the waveguide is reported in which the pump signal is the lowest-order mode of the waveguide, and the generated second harmonic signal corresponds to the third propagation mode of the waveguide. Using a polymer waveguide core, having $\chi^{(2)}$ ~100 pm/V, we predict a conversion efficiency of approximately 90% after a propagation distance of 2 mm, using peak pump intensities inside the core of the waveguide of 1.35 GW/cm$^2$. If the waveguide core contains polymer layers with different glass transition temperatures, the layers can be poled independently to maximize the overlap integral, and similar pump depletions may be achieved over a distance of approximately 500 microns.

## 1. Introduction

Optical second harmonic generation (SHG) has been intensively studied and demonstrated both in bulk and waveguide configurations. For strong SHG, the interacting fields at the fundamental and second harmonic (SH) wavelengths should be well confined and overlapped inside the structure. Usually, the length of any device devoted to obtain highly efficient SHG is limited by the phase-mismatch between the waves involved in the nonlinear process. When the phase-matching condition is achieved, conversion efficiencies can be large by simply increasing the interaction length.

If the dispersion is not too strong, high conversion efficiencies can typically be achieved by resorting to the quasi-phase-matching (QPM) technique [1], where a periodic modulation of the second order, nonlinear susceptibility tensor compensates the phase-mismatch between waves traveling at different phase-velocities. It was recently demonstrated [2,3] that enhanced SHG may be obtained in 1-dimensional photonic crystal devices by tuning both the fundamental and SH signals near their respective band edges: geometrical and material dispersions may be combined and exploited to tune the effective phase matching conditions near the band edge, where the density of mode is high for both fields [2].

In this paper we show that it is possible to achieve phase-matched SHG by guiding the pump field energy in an ordinary, quadratic nonlinear material, sandwiched between two Bragg-like mirrors constituted by one-dimensional, sub-wavelength thick, multilayer structures. The operation of such a waveguide in the linear regime is based on Bragg confinement rather than total internal reflection (TIR), or step-index guiding, and allows low-loss and single-mode propagation of the electromagnetic field, which is trapped into the high-nonlinear-coefficient region. This wave-guiding mechanism, first discussed by Yeh and Yariv [4], is exploited in Bragg reflection waveguides (BRW), as well as in micro-structured photonic crystal fibers (PCF), or hollow fibers [5,6], which typically display two-dimensional photonic band gap (PBG) confinement of the light into the fiber core. Recently, a photonic-crystal Mach-Zehnder modulator based on this wave-guiding mechanism, surrounded by metal electrodes, was proposed [7]. The device showed exceptionally high confinement and low-loss properties, effectively reducing electrode separation by at least a factor of two. BRW can also support nonlinear guided wave propagation, with a third-order nonlinear medium either in the core region [8], or in the high-index films of the cladding [9]. In this paper, we suggest the use of photonic-crystals-assisted waveguides to achieve highly-efficient SHG by exploiting field confinement and effective phase matching conditions between the pump and the generated second harmonic signal in the core region, a low-index medium such as a polymer, for example, possessing a quadratic nonlinear coefficient.

The linear propagation properties of this particular type of waveguide strongly depend on the multilayer-cladding design, whose geometry and refractive indices uniquely determine transverse photonic band gaps (PBG) in the linear transmission spectrum of the waveguide [7]. Here, we use the beam propagation method (BPM) to analyze a structure of finite transverse and longitudinal extent. A numerical study of the geometrical properties of the structure and effective refractive index dispersion reveals that nearly exact phase matching can be achieved between the fundamental and the SH signals. Thus, we predict highly efficient SHG by simply molding the effective index properties of the cladding, without the need of introducing a grating inside the core [e.g. 10]. In particular, our calculations show that a 1.35 GW/cm$^2$ pump signal may be almost completely depleted after propagating approximately 2 mm inside the waveguide.

In waveguide geometries, phase matching can be achieved by using higher order modes at the SH frequency to compensate for material dispersion. Using higher order modes to achieve phase matching degrades the overlap of the fundamental and SH fields. Khurgin [11] suggested that by appropriately inverting the nonlinear domains along the transverse coordinate, the overlap integral, and hence conversion efficiency, could be further optimized. In ref. [12] a waveguide core consisting of two nonlinear polymer layers having different glass transition temperatures, allowed for sequential poling of the individual layers composing the core. These inverted domains in the transverse direction provide a large overlap integral for the fundamental and SH fields. As we will show in the following, photonic-crystal-assisted waveguides in combination with modal phase matching and sequential poling can provide extremely large conversion efficiencies.

## 2. Modeling technique

We start our analysis by considering the standard wave equation:

$$\nabla \times (\nabla \times \mathbf{E}(x,z,t)) + \frac{n^2(x)}{c^2}\frac{\partial^2 \mathbf{E}(x,z,t)}{\partial t^2} = -\mu_0 \frac{\partial^2 \mathbf{P}_{NL}(x,z,t)}{\partial t^2} \quad (1)$$

where the refractive index is assumed to be discontinuities along the *x*-axis, and $\mathbf{P}_{NL}(x,z,t) = \varepsilon_0 \chi^{(2)} \mathbf{E}^2(x,z,t)$. For simplicity we consider a TE-polarized field, so that the electric field is aligned along the *y*-axis and Eq.(1) reduces to a scalar problem. We express the electric field as a superposition of monochromatic, forward-propagating waves as follows:

$$E_y(x,z,t) = E_F(x,z)e^{-i(\omega_F t - k_F z)} + E_{SH}(x,z)e^{-i(2\omega_F t - k_{SH} z)} + c.c. \quad (2)$$

In the above expression $k_F$ and $k_{SH}$ are the reference wave-numbers for the fundamental and SH waves, respectively. For convenience, we choose the low-index core layer as the reference medium of the structure, for both the fundamental and SH wavelengths. Substituting expression (2) in the wave equation (1), and after separating the terms oscillating at $\omega_F$ from those oscillating at $2\omega_F$, we obtain the following scalar system of coupled wave equations:

$$\begin{cases} \dfrac{\partial^2 E_F}{\partial x^2} + \dfrac{\partial^2 E_F}{\partial z^2} + 2ik_F \dfrac{\partial E_F}{\partial z} + k_F^2 \left[\left(\dfrac{n_F}{n_{ref,F}}\right)^2 - 1\right] E_F = -2\dfrac{\omega_F^2}{c^2}\chi^{(2)} E_F^* E_{SH} e^{i\Delta k z} \\ \dfrac{\partial^2 E_{SH}}{\partial x^2} + \dfrac{\partial^2 E_{SH}}{\partial z^2} + 2ik_{SH} \dfrac{\partial E_{SH}}{\partial z} + k_{SH}^2 \left[\left(\dfrac{n_{SH}}{n_{ref,SH}}\right)^2 - 1\right] E_{SH} = -\dfrac{4\omega_F^2}{c^2}\chi^{(2)} E_F^2 e^{-i\Delta k z} \end{cases} \quad (3)$$

where the subscripts *F* and *SH* stand for fundamental and second harmonic, $\chi^{(2)}$ is the second order nonlinear coefficient, and $\Delta k = k_{SH} - 2k_F$ is the phase mismatch imposed by the two different reference refractive-indexes $n_{ref,F}$ and $n_{ref,SH}$, which are the low-indices of the core medium at the fundamental and SH frequencies, respectively.

So far the only assumption we have made is that the fields are monochromatic and TE-polarized. In a waveguide environment, the complexity of equations (3) may be reduced by assuming that the refractive index has discontinuities only along the transverse direction *x*, and that it is uniform along the direction of light propagation *z*. This allows us to neglect the second order longitudinal derivatives of the fields, and to solve the parabolic or Fresnel problem, defined as follows

$$\begin{cases} 2ik_F \dfrac{\partial \tilde{E}_F}{\partial z} = -\dfrac{\partial^2 \tilde{E}_F}{\partial x^2} - k_F^2 \left[ \left( \dfrac{n_F}{n_{ref,F}} \right)^2 - 1 \right] \tilde{E}_F - 2\dfrac{\omega_F^2}{c^2} \chi^{(2)} \tilde{E}_F^* \tilde{E}_{SH} e^{j\Delta k z} \\ 2ik_{SH} \dfrac{\partial \tilde{E}_{SH}}{\partial z} = -\dfrac{\partial^2 \tilde{E}_{SH}}{\partial x^2} - k_{SH}^2 \left[ \left( \dfrac{n_{SH}}{n_{ref,SH}} \right)^2 - 1 \right] \tilde{E}_{SH} - \dfrac{4\omega_F^2}{c^2} \chi^{(2)} \tilde{E}_F^2 e^{-j\Delta k z} \end{cases} \quad (4)$$

The system (4) is solved using the split-step Fast-Fourier-Transform beam propagation method (FFT-BPM) [13-14], which has been extended to include the simultaneous propagation of two fields, the fundamental and the SH, coupled via the quadratic nonlinear susceptibility. Using this method, we are also able to fully describe the dispersion properties of the structure that we analyze. The procedure used to determine the eigenvalues and eigenmodes, which is described in [14], is based on the correlation function between the electric field propagating through the waveguide, and the source, or input electric field. This technique is applied by considering the linear behavior of the waveguide, thus separating the fundamental and the second harmonic equations of motion by removing the nonlinear coupling terms in (4). The eigenvalues of the waveguide correspond to the maxima of the Fourier transform of the above mentioned correlation. Once the eigenvalues $\beta_n$ are known, the effective index is given by

$$n_{eff,F/SH} = \dfrac{c}{\omega_{F/SH}} \left( k_{F/SH} + \beta_n \right), \quad (5)$$

while the eigenmodes are evaluated by $\dfrac{1}{L} \int_0^L \tilde{E}(x,z) w(z) e^{i\beta_n z} dz$ [13], where $w(z)$ is a window function.

## 3. Results and discussion

The basic structure we analyze is schematized in Fig.1: dark regions correspond to the high-refractive-index medium. The period $\Lambda$ of the lateral Bragg mirrors and the index contrast $\Delta n = |n_2 - n_1|$ define the width and the spectral position of the band gap in the longitudinal linear transmission of the waveguide. As pointed out in Ref. [7], the width of the transmission bands increases when the index contrast in the lateral cladding regions is larger. Assuming material and scattering losses due to lattice imperfections may be neglected, propagation losses in the waveguide decrease by increasing the number of periods. Moreover, it was shown [7] that the waveguide performance, in terms of losses over transmission and reduction of design complexity, is good when only two media are considered, instead of three, as in the most general case. The structure we propose is constituted by a core layer (refractive index $n_c(\omega_F) = 1.6$ at the fundamental wavelength, typical for polymers) 2 μm thick. The claddings are composed of periodic arrangements of semiconductor layers (refractive index $n_1(\omega_F) = 3.2$) and polymeric layers; the period is $\Lambda = 300$ nm, and each layer is 150 nm thick. The number of periods is N=6, and the total thickness of the structure is less than 4 μm. We assume that the refractive index of the polymer at the SH wavelength is $n_c(2\omega_F) = 1.6283$, consistent with a chromatic dispersion of approximately 1.8%, a typical but by no means unique range for ordinary polymers.

For such a waveguide, the longitudinal transmission spectrum, i.e. propagating along z in Fig. 1, is shown in Fig. 2 for a wide range of wavelengths. For our purposes it is enough for the input signal to be propagated for approximately 1 cm. For this analysis we excite the waveguide with a transverse field profile that corresponds to a cosine function within the guiding layer. This represents a good approximation of the shape of the lowest-order propagating mode [7]. For the linear transmission study we use this cosine-shaped input for the entire spectrum that we consider, including waves involved in the nonlinear process. In doing so we are able to fully describe the transmission and dispersion properties of the waveguide, including higher order modes. Referring to Fig.2, interacting signals are chosen within the two different longitudinal pass-bands shown, and they propagate well-confined by the lateral mirrors as the high transmission suggests. Propagation losses are exceptionally small, predicted to be less than 0.005 dB/cm for the fundamental signal [7], and approximately an order of magnitude smaller for the SH signal. Some of this loss is due to the fact that the input field (the cosine function) is a good approximation but does not perfectly match the eigenmode profile.

Then we separately study the modal dispersion of this waveguide for the fundamental wavelength $\lambda_F = 1064$ nm, and SH wavelength $\lambda_{SH} = 532$ nm. In Fig. 3 the modal powers are plotted on a logarithmic scale as functions of the effective refractive index for the fundamental (solid line) and SH (dashed line) wavelengths, respectively. These curves show that the waveguide supports multimode propagation, but the lowest-order mode carries by far (several orders of magnitude) the largest amount of modal power. Moreover, it is evident that the waveguide produces a phase-matching condition between the TE$_0$ mode at $\lambda_F = 1064$ nm, the TE$_2$ mode at the SH wavelength $\lambda_{SH} = 532$ nm, so that the third peak of the SH modal power spectrum (dashed line in Fig.3), falls in almost precisely the same spectral position of the TE$_0$ mode at the fundamental wavelength (first peak in the continuous line). The corresponding effective refractive index value is $n_{eff} = 1.5795$. An important feature of this waveguide is that the magnitude of the effective refractive index for the guided mode can be smaller than the individual refractive indices in each layer. This is due to the Bragg nature of the guiding system, so that the lateral, multilayer stacks act as mirrors for the light inside the core region, thus yielding an effective refractive index smaller than the index in the core region. This reduction of the effective refractive index was pointed out in reference [2] for one-dimensional PBG structures, but for excitations normal to the layer interfaces, at frequencies inside photonic band gap. The lower effective refractive index comes about thanks to the onset of anomalous geometrical dispersion, which compensates natural material dispersion. Therefore, using the geometry external to the guiding layer one is able to compensate for material *and* modal dispersions to the extent that nearly exact phase matching can be achieved.

The transverse profiles of the two interacting modes are reported in Fig.4. The phase-matching in this waveguide is also confirmed by examining the dispersion curves (Fig.5) of the effective refractive index for the $TE_0$ mode near $\lambda_F = 1064$ nm, and the $TE_2$ mode near $\lambda_{SH} = 532$ nm. We note that these curves are obtained using the method outlined in section 2 above. However, an independent calculation of the dispersive properties of the guide using an analytical model based on mode matching at each interface yields almost identical results.

Using a pump intensity peak $I_\omega$=1.35 GW/cm$^2$, the conversion efficiency $\eta = P_{2\omega}/P_\omega$ is about 90% after a propagation distance of 2mm, as depicted in Fig.6, where the cross-sectional power of the fundamental and SH fields are plotted as a function of the propagation distance z.

The structure that we have studied is by no means unique, and it is optimized for a pump wavelength $\lambda_F = 1064$ nm. However, geometrical dispersion makes it possible to shift the operating wavelength of the waveguide, thus shifting the phase-matching condition, by simply varying the cladding layers thicknesses. In general, the phase-matching wavelength is red-shifted when the width of the layers increases. For instance, we found that for layers 180 nm thick the phase-matched wavelength is 1.085 μm. Of course, when the periodicity of each multilayer stack is changed (and they may be changed independently), particular attention has to be devoted to changes in the longitudinal transmission spectrum.

An examination of Fig. 4, which shows the transverse field distributions at the fundamental and second harmonic frequencies, suggests that it may be possible to further enhance the conversion efficiency by appropriately inverting the nonlinear domains along the transverse coordinate to maximize field overlap [11,12]. In Fig. 7 we compare the conversion efficiency for three different structures: (i) one coherence length of core material; (ii) a photonic crystal assisted waveguide with an unprocessed core 500 μm-long; (iii) same as (ii), but with a transversely poled core. The result of the transverse poling is a dramatic improvement of the second harmonic conversion process: a 1.35 GW/cm$^2$ input pump beam may become nearly depleted after a propagation distance of only 500 microns. Alternatively, the same 2mm-long waveguide may almost completely deplete an 80 MW/cm$^2$ pump signal.

## Conclusions

We have shown second harmonic generation in a photonic-crystal-assisted waveguide in a mode dispersion phase matching configuration. The high Bragg-like confinement of the interacting waves provided by the periodic structures in the cladding regions makes this waveguide particularly interesting for highly-efficient nonlinear processes in low-refractive index media, such as polymers. The conversion efficiency for a waveguide with a uniformly poled polymer core waveguide is about 90% using a 1.35 GW/cm$^2$ input pump power and a 2 mm long device. For waveguide cores containing two or more polymer layers with different glass transition temperatures, the application of transverse poling to maximize field overlap leads to a device capable of depleting a 1.35 GW/cm$^2$ pump in only 500 μm.

**Acknowledgements**

D. de Ceglia thanks the European Research Office (United States Army), research contract n. N62558-05-P-0486 and FIRB 2001 research contract RBAU01XEEM for financial supports.

**Figure Captions**

**Fig.1:** Sketch of the photonic-crystal assisted waveguide. The refractive index of the core region (grey) is smaller than the average index of the periodic cladding (black-white layers).

**Fig.2:** Transmissivity of the waveguide. Both the fundamental (around 1.064 μm) and the generated second harmonic ( around 0.532 μm) signals are transmitted through the waveguide.

**Fig.3:** Modal power versus effective refractive index for the fundamental λ=1.064 μm (solid curve) and second harmonic (dotted curve) wavelengths in logarithmic scale. Each peak is the power carried by a guided mode. Lower order modes correspond to higher refractive index positions.

**Fig.4:** Profiles of the mode intensities at the fundamental (continuous line) and second harmonic (dashed line) wavelengths. The fundamental signal is a $TE_0$ mode, the phase-matched SH mode is a $TE_2$. The alternation of the sign of $\chi^{(2)}$ refers to transverse poling of the structure, discussed in section three, to further improve the overlap integral and overall conversion efficiencies.

**Fig.5:** Dispersion curves for the $TE_0$ (continuous) and $TE_2$ (dotted) modes. The wavelength axis for the $TE_2$ mode is multiplied by a factor two. The open circle denotes the phase-matching wavelength, $\lambda=1.064$ μm.

**Fig.6:** Pump (solid line) and SH (dashed line) powers versus interaction length. The power of the input pump signal is 1.35 GW/cm$^2$. The pump is depleted after 2 mm of propagation.

**Fig.7:** SHG efficiency versus pump power for three waveguides. The dotted line is relative to a one-coherence-length-long waveguide, constituted by a polymeric core surrounded by air. The dashed line corresponds to the photonic-crystal-assisted, 500μm-long waveguide. The solid curve is the efficiency obtained by transversally poling the photonic-crystal assisted waveguide.

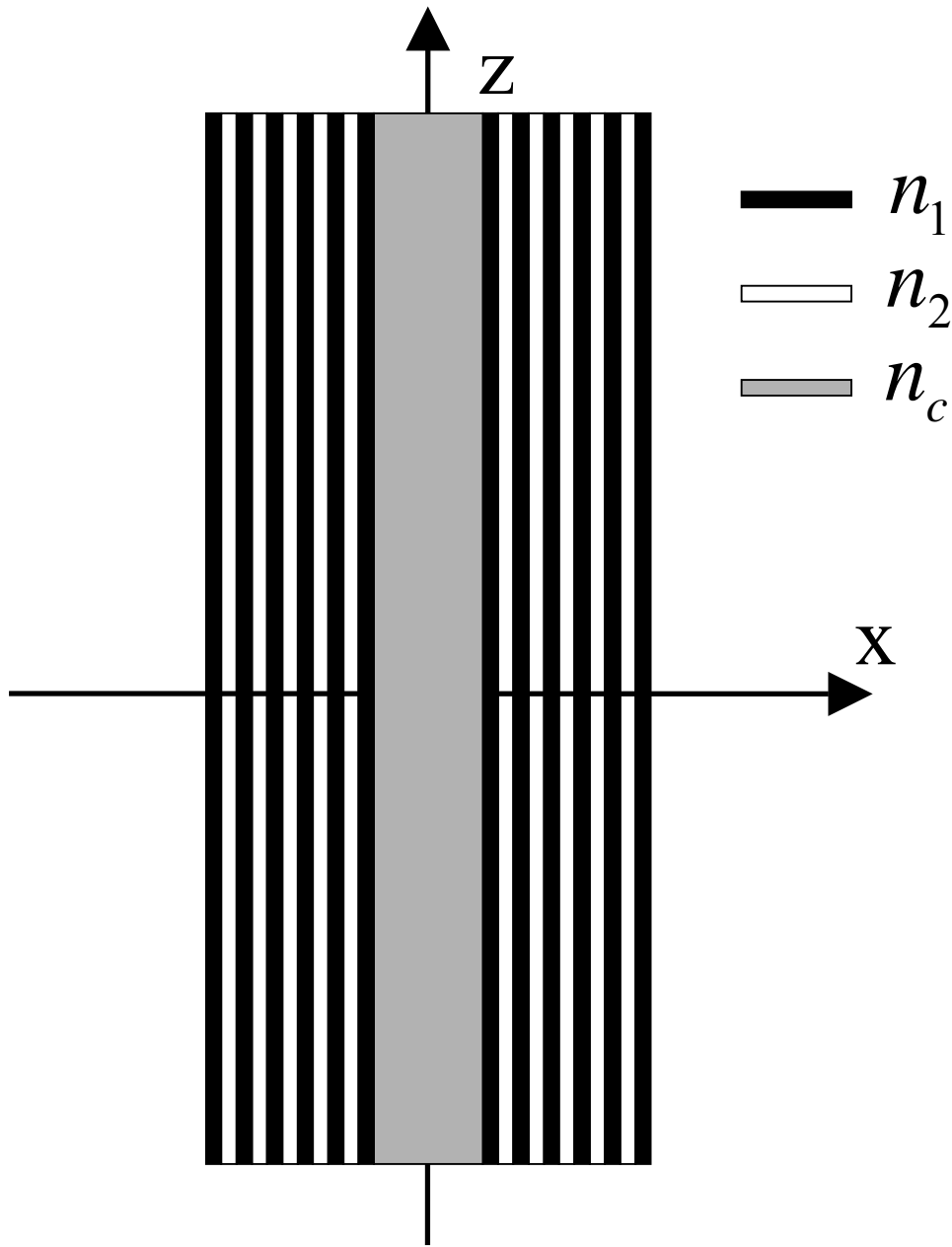

Fig.2

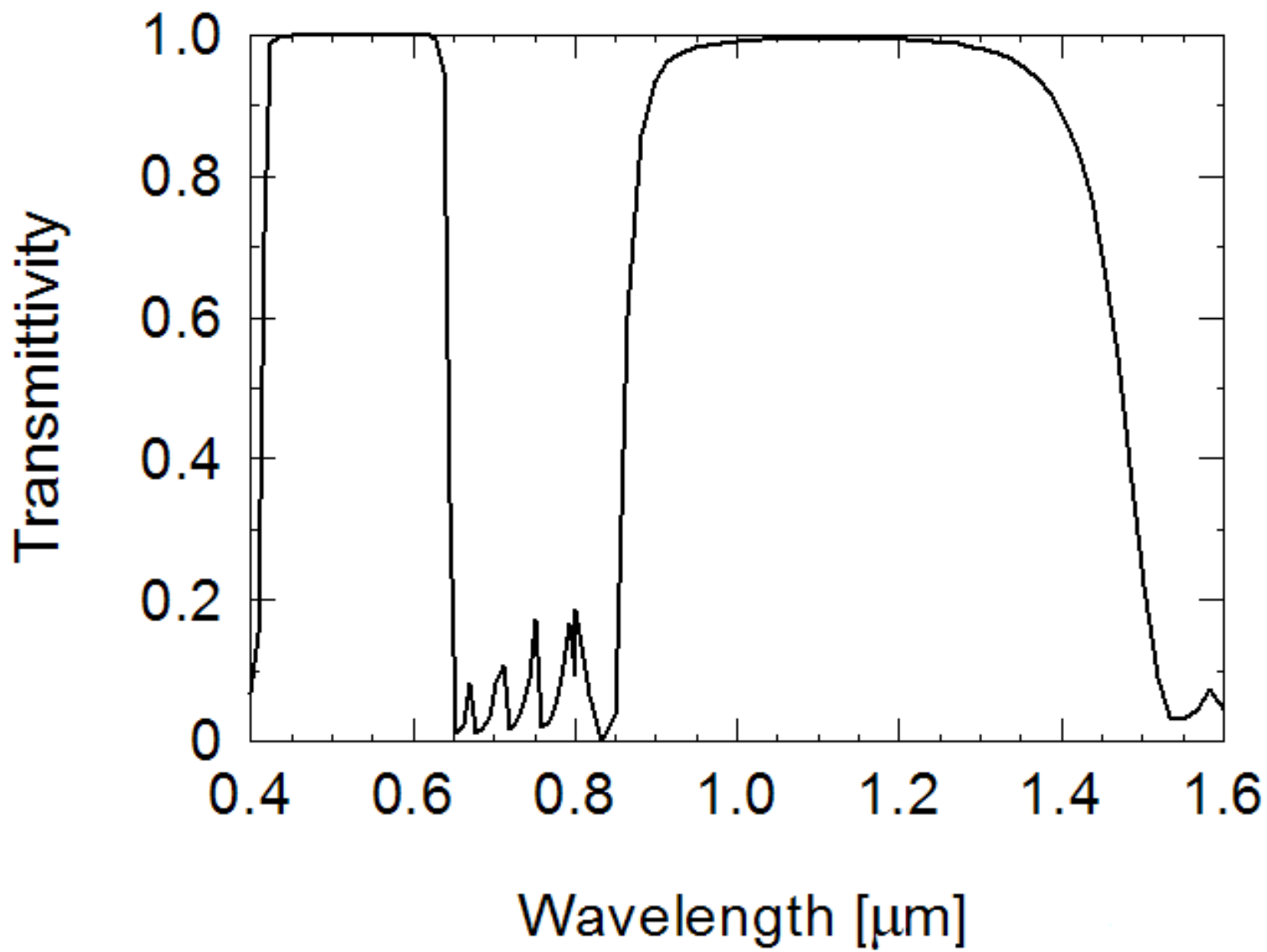

Fig.3

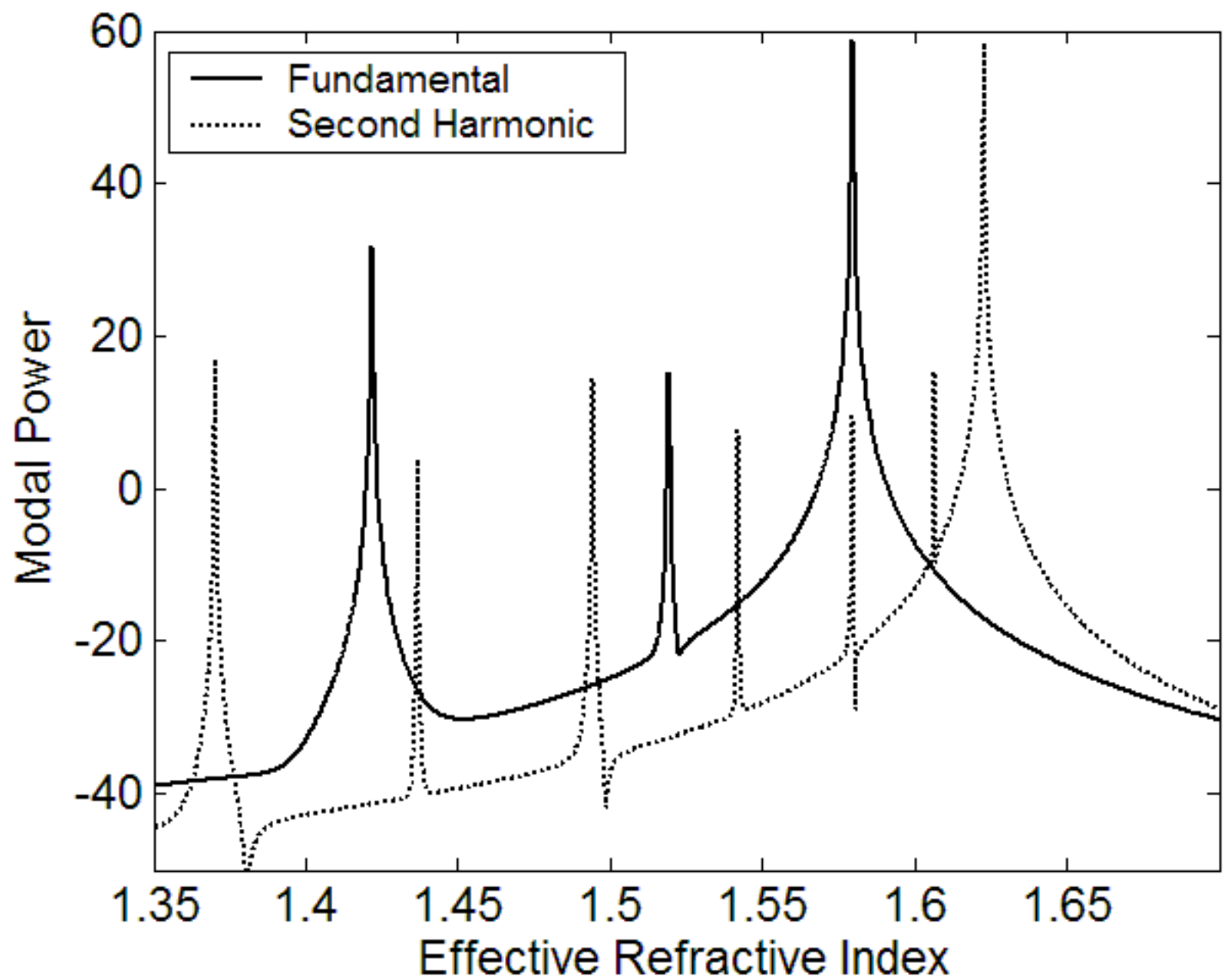

Fig.4

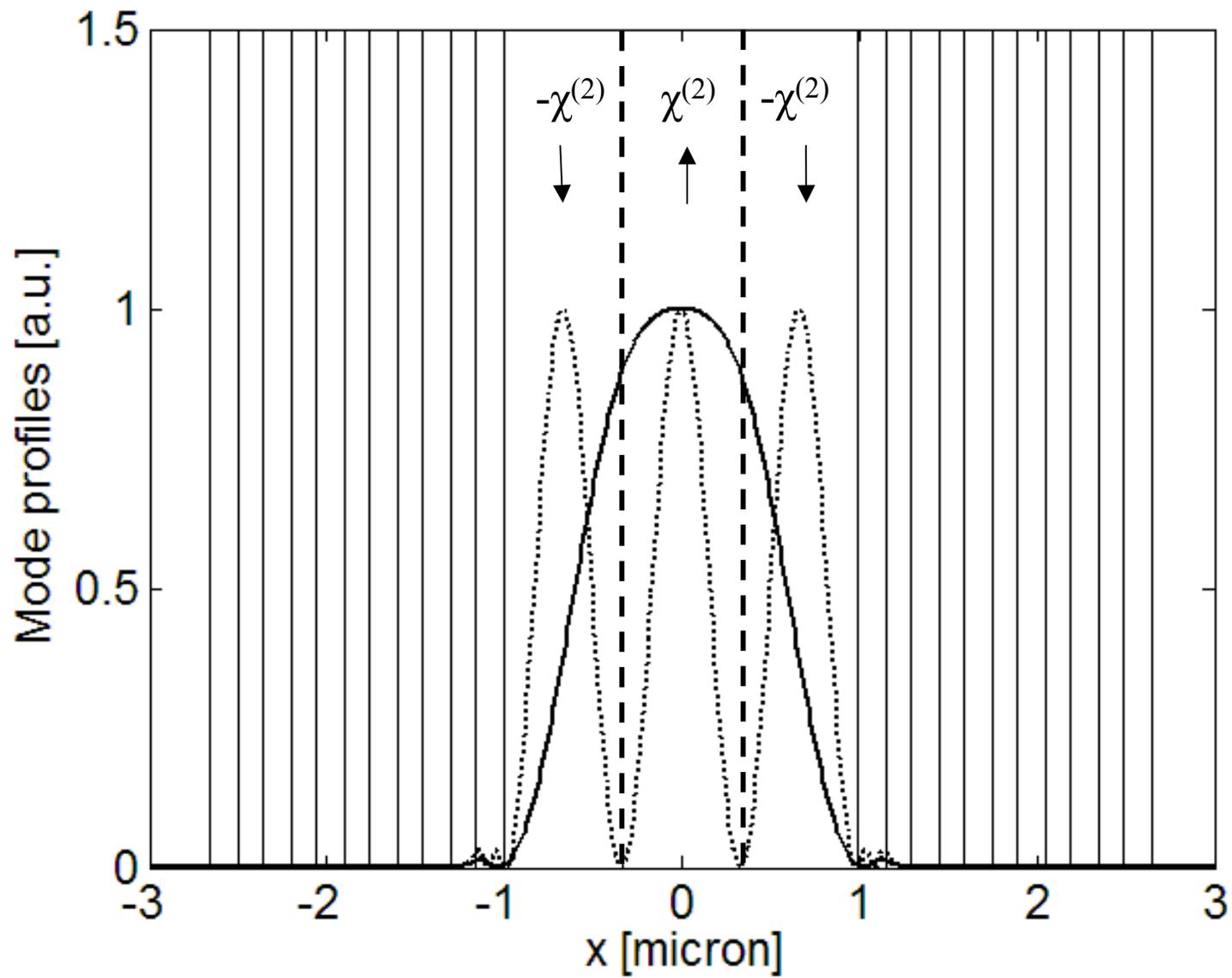

Fig.5

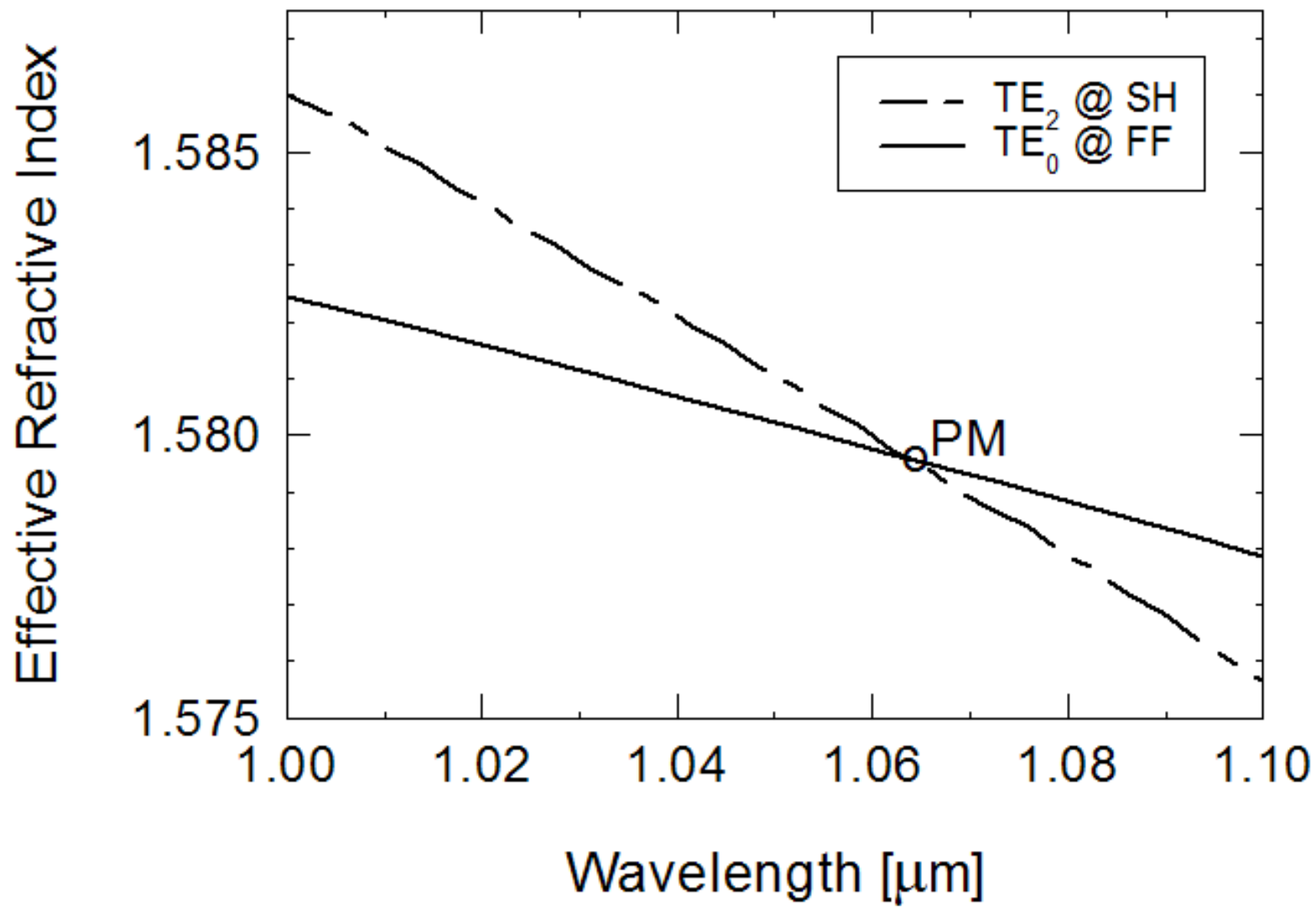

Fig.6

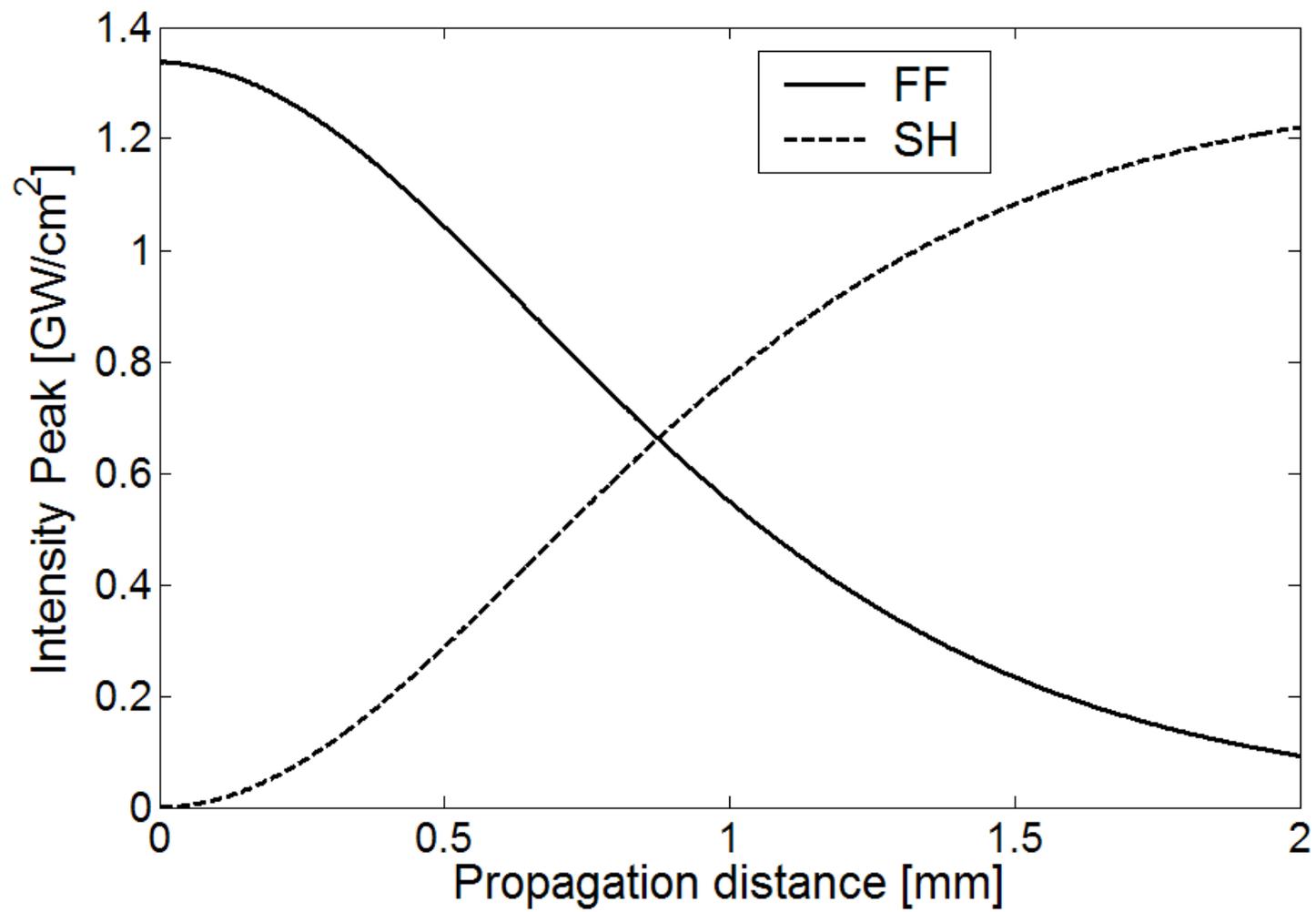

Fig.7

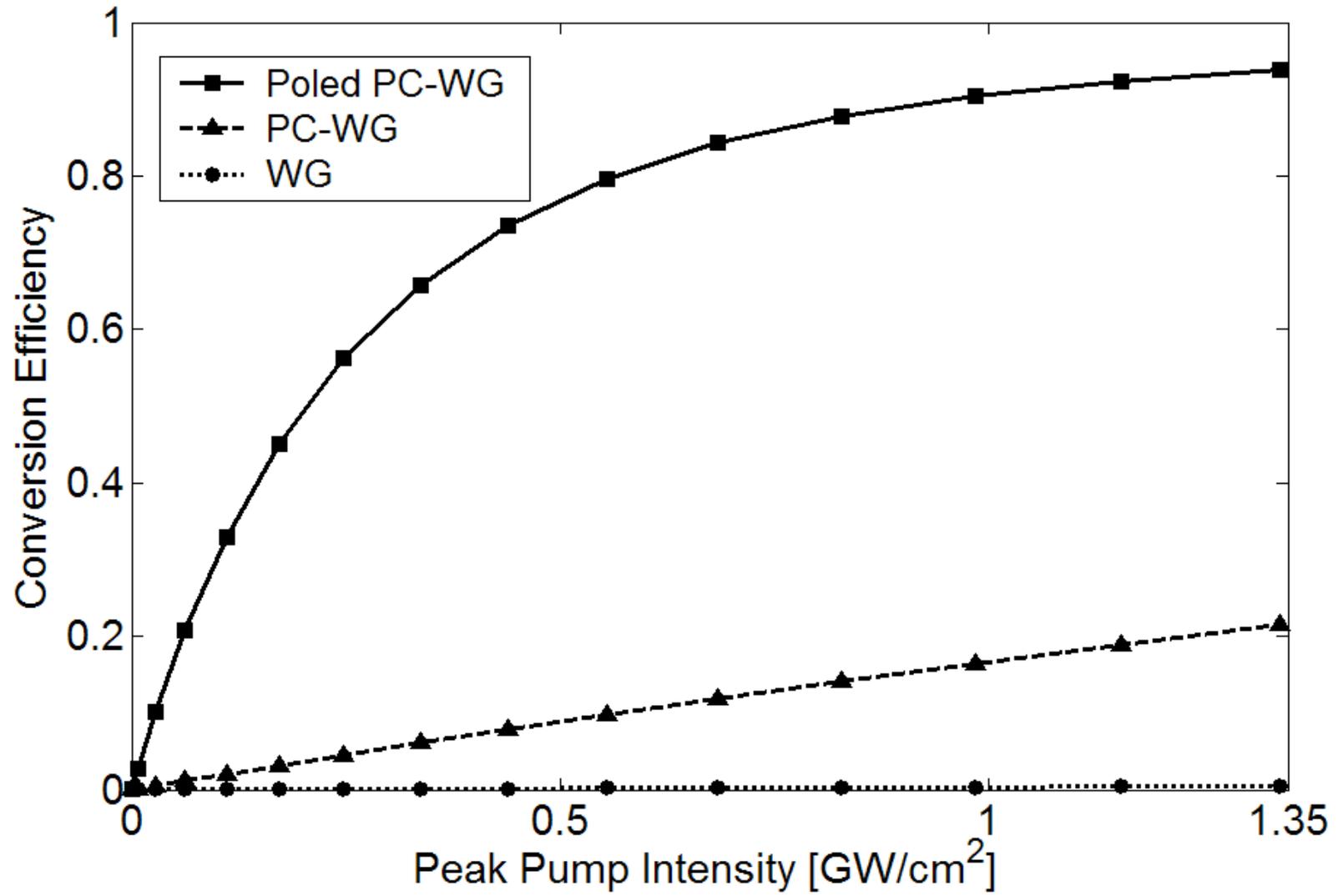